\documentclass[a4paper]{article}
\usepackage{graphicx}
\usepackage[compress]{cite}
\usepackage{braket}
\usepackage{bm}
\usepackage{comment}
\usepackage{here}
\usepackage{a4wide}
\usepackage{authblk}
\begin{document}
\def\Journal#1#2#3#4{{#1} {\bf #2}, #3 (#4)}
\def\AHEP{Advances in High Energy Physics.} 	
\def\ARNPS{Annu. Rev. Nucl. Part. Sci.} 
\def\AandA{Astron. Astrophys.} 
\def\ANP{Ann. Phys.}
\def\APJ{Astrophys. J.}
\def\APJS{Astrophys. J. Suppl}
\def\COMR{Comptes Rendues}
\def\CQG{Class. Quantum Grav.}
\def\CPC{Chin. Phys. C}
\def\EPJC{Eur. Phys. J. C}
\def\EPL{EPL}
\def\IJMPA{Int. J. Mod. Phys. A}
\def\IJMPE{Int. J. Mod. Phys. E}
\def\JCAP{J. Cosmol. Astropart. Phys.}
\def\JHEP{J. High Energy Phys.}
\def\JETPL{JETP. Lett.}
\def\JETPUSSR{JETP (USSR)}
\def\JPG{J. Phys. G} 
\def\JPCS{J. Phys. Conf. Ser.} 
\def\JPGNP{J. Phys. G: Nucl. Part. Phys.} 
\def\MPLA{Mod. Phys. Lett. A}
\def\NIMA{Nucl. Instrum. Meth. A.}
\def\NATU{Nature}
\def\NCA{Nuovo Cimento}
\def\NJP{New. J. Phys.}
\def\NPB{Nucl. Phys. B}
\def\NPBOLD{Nucl. Phys.}
\def\NPBSUPPL{Nucl. Phys. B. Proc. Suppl.}
\def\PL{Phys. Lett.}
\def\PLB{{Phys. Lett.} B}
\def\PMCA{PMC Phys. A}
\def\PREP{Phys. Rep.}
\def\PPNP{Prog. Part. Nucl. Phys.}
\def\PLBOLD{Phys. Lett.}
\def\PAN{Phys. Atom. Nucl.}
\def\PRL{Phys. Rev. Lett.}
\def\PRD{Phys. Rev. D}
\def\PRC{Phys. Rev. C}
\def\PR{Phys. Rev.}
\def\PTP{Prog. Theor. Phys.}
\def\PTEP{Prog. Theor. Exp. Phys.}
\def\RMP{Rev. Mod. Phys.}
\def\RPP{Rep. Prog. Phys.}
\def\SJNP{Sov. J. Nucl. Phys.}
\def\SCIENCE{Science}
\def\TNYAS{Trans. New York Acad. Sci.}
\def\ZETP{Zh. Eksp. Teor. Piz.}
\def\ZFPH{Z. fur Physik}
\def\ZPC{Z. Phys. C}
\title{Is magic texture for Majorana neutrino immanent in Dirac nature?}
\author[1]{Yuta Hyodo \footnote{Corresponding author} \footnote{2ctad004@mail.u-tokai.ac.jp}}
\author[2]{Teruyuki Kitabayashi\footnote{teruyuki@tokai-u.jp}}
\affil[1]{Graduate School of Science and Technology, Tokai University, 4-1-1 Kitakaname, Hiratsuka, Kanagawa 259-1292, Japan}
\affil[2]{Department of Physics, Tokai University, 4-1-1 Kitakaname, Hiratsuka, Kanagawa 259-1292, Japan}
\date{}
\maketitle
\begin{abstract}
The magic textures are successful candidates of the correct texture for the Majorana neutrinos. In this study, we show that some types of magic texture for Majorana neutrinos approximately immanent in the flavor mass matrix for Dirac neutrinos. In addition, it turned out that the normal mass ordering of the Dirac neutrino masses is slightly preferable to the inverted mass ordering in the context of the magic textures.
\end{abstract}

\section{Introduction\label{section:introduction}}
Are neutrinos Majorana or Dirac particles? This statistical nature of neutrinos is one of the unresolved issues in neutrino physics.

The flavor neutrino mass matrix for Majorana-type neutrinos is symmetric. However the flavor neutrino mass matrix for Dirac-type neutrinos is not symmetric. Therefore, the texture of the Dirac flavor neutrino mass matrix has a more complicated structure than the Majorana flavor neutrino mass matrix. It is relatively difficult to find symmetry for the Dirac neutrino mass matrix. It would make sense to explore the Majorana flavor mass matrix immanent in the Dirac flavor mass matrix. This immanence provides a clue to solve the problem about the statistical nature of neutrinos. 

In this paper, we focus on magic texture, which is one of the successful textures of Majorana flavor neutrino mass matrix\cite{Harrison2004PLB,Lam2006PLB,Gautam2016PRD,Yang2021arXiv,Channey2019JPGNP,Verma2020JPGNP,Hyodo2021PTEP, Verma2022arXiv}. We investigated the relationship between Dirac flavor neutrino mass and Majorana flavor neutrino mass matrices. 

There are several types of magic textures reproduced in the results of neutrino oscillation experiments. The first discovered traditional magic texture of Majorana flavor neutrino mass matrix is parameterized as follows\cite{Harrison2002PLB}:
\begin{eqnarray}
&& \qquad \quad
\left(
\begin{array}{ccc}
a & b & c \\
b & d & a+c-d \\
c & a+c-d & b-c+d
\end{array}
\right)
\begin{array}{c}
\leftarrow a+b+c \\
\leftarrow a+b+c \\
\leftarrow a+b+c .\\
\end{array}
\label{Eq:magicTextureHarrison}
\end{eqnarray}
The traditional magic texture is invariant under $Z_2$ symmetry. Indeed, the traditional magic texture is independent of the neutrino mass eigenvalues $(m_1,m_2,m_3)$ and is related only to the mixing matrix\cite{Lam2006PRD}.

The relationship between the texture of the flavor neutrino mass matrix for Dirac-type neutrinos and magic texture has been studied in Ref.\cite{Hyodo2020IJMPA}. We explain the difference between the previous study and this study. There is no discussion about magic texture for Majorana neutrino immanent in texture of Dirac flavor mass matrix in the previous study\cite{Hyodo2020IJMPA}. Meanwhile, this immanence is the main topic in this paper. Additionally, although the elements of the Dirac flavor neutrino mass matrix are complex, it is assumed that the elements of the mass matrix are real in Ref.\cite{Hyodo2020IJMPA}. In this paper, the elements of the mass matrix are complex.

Other than magic texture, many possible textures of flavor mass matrix for Majorana-type have been proposed in the literature, such as tri-bi maximal texture \cite{Harrison2002PLB,Xing2002PLB,Harrison2002PLB2,Kitabayashi2007PRD}, texture zeros \cite{Berger2001PRD,Frampton2002PLB,Xing2002PLB530,Xing2002PLB539,Kageyama2002PLB,Xing2004PRD,Grimus2004EPJC,Low2004PRD,Low2005PRD,Grimus2005JPG,Dev2007PRD,Xing2009PLB,Fritzsch2011JHEP,Kumar2011PRD,Dev2011PLB,Araki2012JHEP,Ludle2012NPB,Lashin2012PRD,Deepthi2012EPJC,Meloni2013NPB,Meloni2014PRD,Dev2014PRD,Felipe2014NPB,Ludl2014JHEP,Cebola2015PRD,Gautam2015PRD,Dev2015EPJC,Kitabayashi2016PRD1,Zhou2016CPC,Singh2016PTEP,Bora2017PRD,Barreiros2018PRD,Kitabayashi2018PRD,Barreiros2019JHEP,Capozzi2020PRD,Singh2020EPL,Barreiros2020,Kitabayashi2020PRD,Kitabayashi2017IJMPA,Kitabayashi2017IJMPA2,Kitabayashi2019IJMPA}, $\mu-\tau$ symmetric texture \cite{Fukuyama1997,Lam2001PLB,Ma2001PRL,Balaji2001PLB,Koide2002PRD,Kitabayashi2003PRD,Koide2004PRD,Aizawa2004PRD,Ghosal2004MPLA,Mohapatra2005PRD,Koide2005PLB,Kitabayashi2005PLB,Haba206PRD,Xing2006PLB,Ahn2006PRD,Joshipura2008EPJC,Gomez-Izquierdo2010PRD,He2001PRD,He2012PRD,Gomez-Izquierdo2017EPJC,Fukuyama2017PTEP,Kitabayashi2016IJMPA,Kitabayashi2016PRD,Bao2021arXiv,Garces2018JHEP,JuanCarlos2019JHEP}, and textures under discrete symmetries, e.g., $A_n$ and $S_n$\cite{Altarelli2010PMP}. 

The remainder of the paper is organized as follows. In Sect.\ref{section:Majorana_magic}, we review magic textures of neutrino mass matrix for Majorana-type. In Sect.\ref{section:Dirac_magic_immanent}, we estimate Majorana-type neutrino mass matrix immanent in Dirac-type neutrino mass matrix in terms of magic textures. Sect.\ref{section:summary} presents the summary.
\section{Majorana flavor neutrino mass matrix and magic textures\label{section:Majorana_magic}}
We review magic textures of mass matrix for Majorana-type neutrinos\cite{Harrison2004PLB,Lam2006PLB,Hyodo2021PTEP}.

Since the Majorana neutrino flavor mass matrix is symmetric, there are five independent sums, $\{S_{\rm M1}, S_{\rm M2}, S_{\rm M3}, S_{\rm M4}, S_{\rm M5}\}$, schematically:
\begin{eqnarray}
M_{\rm M}=
\left(
\begin{array}{ccc}
 \ a\  & \ b \ &  \ c \ \\
 b & d & e \\
 c & e & f \\
\end{array}
\right) 
\begin{array}{c}
\leftarrow S_{\rm M1}\\
\leftarrow S_{\rm M2}\\
\leftarrow S_{\rm M3}\\
\end{array}
\nonumber \\
\begin{array}{ccccccc}
&\nearrow & ~\uparrow & ~\uparrow & ~\uparrow & \nwarrow & \\
S_{\rm M5}&&S_{\rm M1} & S_{\rm M2}& S_{\rm M3} &&S_{\rm M4}\\
\end{array}
\label{Eq:MajoranaFlavorNeutrinoMassMatrix}
\end{eqnarray}
where
\begin{eqnarray}
S_{\rm M1}= a+b+c, \quad
S_{\rm M2}= b+d+e, \quad
S_{\rm M3}= c+e+f
\label{Eq:SM1SM2SM3}
\end{eqnarray}
for $i$th raw ($i$th column) and
\begin{eqnarray}
S_{\rm M4}= a+d+f, \quad
S_{\rm M5}= 2c + d,
\label{Eq:SM4SM5}
\end{eqnarray}
for diagonal elements. 

These five sums $\{S_{\rm M1}, S_{\rm M2}, S_{\rm M3}, S_{\rm M4}, S_{\rm M5}\}$, have been used to classify the type of the magic texture of the Majorana neutrino flavor mass matrix \cite{Hyodo2021PTEP}. Based on the success of the first discovered magic texture in Eq.(\ref{Eq:magicTextureHarrison}), three of the five sums being the same in the Majorana neutrino mass matrix are required for other magic textures. Under this requirement, we have $_5C_3=10$ types of magic textures. These ten textures are called type M1, type M2, $\cdots$, and type M10 magic texture with the following definitions\footnote{In this paper, we write type I, type II, $\cdots$, in Ref.\cite{Hyodo2021PTEP} as type M1, M2, $\cdots$.}:
\begin{eqnarray}
&&{\bf M1:} \ S_{\rm M1}=S_{\rm M2}=S_{\rm M3}, \quad
{\bf M2:}  \ S_{\rm M1}=S_{\rm M2}=S_{\rm M4}, \quad 
{\bf M3:} \ S_{\rm M1}=S_{\rm M3}=S_{\rm M4}, \nonumber \\
&&{\bf M4:} \ S_{\rm M2}=S_{\rm M3}=S_{\rm M4}, \quad 
{\bf M5:} \ S_{\rm M1}=S_{\rm M2}=S_{\rm M5} , \quad
{\bf M6:} \ S_{\rm M1}=S_{\rm M3}=S_{\rm M5}, \nonumber \\ 
&&{\bf M7:} \ S_{\rm M2}=S_{\rm M3}=S_{\rm M5}, \quad
{\bf M8:} \ S_{\rm M1}=S_{\rm M4}=S_{\rm M5}, \quad 
{\bf M9:} \ S_{\rm M2}=S_{\rm M4}=S_{\rm M5}, \nonumber \\ 
&&{\bf M10:} \ S_{\rm M3}=S_{\rm M4}=S_{\rm M5} 
\label{Eq:majoranamagictype}
\end{eqnarray}
We have the following ten types of the flavor neutrino mass matrix:
\begin{eqnarray}
M_{\rm M1}=\left(
\begin{array}{ccc}
a & b & c \\
b & e & a+c-e \\
c & a+c-e & b-c+e
\end{array}
\right),
\label{Eq:M_D1}
\end{eqnarray}
\begin{eqnarray}
M_{\rm M2} =\left(
\begin{array}{ccc}
a & b & c \\
b & e & a+c-e \\
c & a+c-e & b+c-e
\end{array}
\right),
\label{Eq:M_D2}
\end{eqnarray}
\begin{eqnarray}
M_{\rm M3} =\left(
\begin{array}{ccc}
a & b & c \\
b & e & a-c+e \\
c & a-c+e & b+c-e
\end{array}
\right),
\label{Eq:M_D3}
\end{eqnarray}
\begin{eqnarray}
M_{\rm M4} =\left(
\begin{array}{ccc}
a & b & c \\
b & e & a-c+e \\
c & a-c+e & b-c+e
\end{array}
\right),
\label{Eq:M_D4}
\end{eqnarray}
\begin{eqnarray}
M_{\rm M5} =\left(
\begin{array}{ccc}
a & b & c \\
b & a+b-c & -b+2c \\
c & -b+2c & i
\end{array}
\right),
\label{Eq:M_D5}
\end{eqnarray}
\begin{eqnarray}
M_{\rm M6} =\left(
\begin{array}{ccc}
a & b & c \\
b & a+b-c & f\\
c & f & a+b-f
\end{array}
\right),
\label{Eq:M_D6}
\end{eqnarray}
\begin{eqnarray}
M_{\rm M7} =\left(
\begin{array}{ccc}
a & b & c \\
b & e & -b+2c \\
c & -b+2c & b-c+e
\end{array}
\right),
\label{Eq:M_D7}
\end{eqnarray}
\begin{eqnarray}
M_{\rm M8} =\left(
\begin{array}{ccc}
a & b & c \\
b & a+b-c & f \\
c & f & -a+2c
\end{array}
\right),
\label{Eq:M_D8}
\end{eqnarray}
\begin{eqnarray}
M_{\rm M9} =\left(
\begin{array}{ccc}
a & b & c \\
b & e & -b+2c \\
c & -b+2c & -a+2c
\end{array}
\right),
\label{Eq:M_D9}
\end{eqnarray}
\begin{eqnarray}
M_{\rm M10} =\left(
\begin{array}{ccc}
a & b & c \\
b & e & a-c+e \\
c & a-c+e & -a+2c
\end{array}
\right).
\label{Eq:M_D10}
\end{eqnarray}

According to the previous study\cite{Harrison2004PLB,Lam2006PLB,Hyodo2021PTEP}, type M1, M4, and M9 are satisfied with observations. Type M1 is a traditional magic texture shown in Eq.(\ref{Eq:magicTextureHarrison}). 
%
\section{Magic texture for Majorana neutrino immanent in Dirac flavor neutrino mass matrix\label{section:Dirac_magic_immanent}}
This study aims to investigate the relationship between the Dirac flavor neutrino mass matrix and magic textures for Majorana-type neutrinos. 

We assume that the mass matrix of charge leptons is diagonal and real. In this case, the Dirac flavor neutrino mass matrix is written as
\begin{eqnarray}
M_{\rm D} = 
\left ( 
\begin{array}{ccc}
a & b & c \\
d & e & f \\
g& h & i \\
\end{array}
\right)
=
 {\rm diag.}(e^{i\phi_e},e^{i\phi_\mu}, e^{i\phi_\tau}) U {\rm diag.}(m_1,m_2, m_3) V^\dag,
\label{Eq:DiracFlavorNeutrinoMassMatrix2}
\end{eqnarray}
where, $\phi_e,\phi_\mu,\phi_\tau \in (-\pi, \pi)$ are phases of the left-handed neutrinos, $m_1,m_2$, $m_3$ are neutrino mass eigenstates, and $U$ is the Pontecorvo-Maki-Nakagawa-Sakata mixing matrix\cite{Pontecorvo1957,Pontecorvo1958,Maki1962PTP,PDG}:
\begin{eqnarray}
U  
=
\left ( 
\begin{array}{ccc}
c_{12}c_{13} & s_{12}c_{13} & s_{13} e^{-i\delta} \\
- s_{12}c_{23} - c_{12}s_{23}s_{13} e^{i\delta} & c_{12}c_{23} - s_{12}s_{23}s_{13}e^{i\delta} & s_{23}c_{13} \\
s_{12}s_{23} - c_{12}c_{23}s_{13}e^{i\delta} & - c_{12}s_{23} - s_{12}c_{23}s_{13}e^{i\delta} & c_{23}c_{13} \\
\end{array}
\right),
\end{eqnarray}
where, $c_{ij}=\cos\theta_{ij}$, $s_{ij}=\sin\theta_{ij}$ ($i,j$=1,2,3), $\theta_{ij}$ is a mixing angle and $\delta$ is Dirac CP-violating phase.

The matrix $V$ is transformation matrix of three right-handed neutrinos\cite{Borgohain2021JPG} which can be parametrized with three rotation angles $\phi_{ij} \in (0, \pi/2)$ $(ij=12, 23,13)$, six phases $\omega_i  \in (-\pi, \pi)$ ($i=1,2,\cdots,5$) and $\delta_V  \in (-\pi, \pi)$. The matrix $V$ can be written by \cite{Borgohain2021JPG,Senjanovic2016PRD}
\begin{eqnarray}
V = {\rm diag.}(e^{i\omega_1}, e^{i\omega_2}, e^{i\omega_3}) \tilde V (\phi_{ij}, \delta_V) {\rm diag.}(e^{i\omega_4}, e^{i\omega_5}, 1) ,
\label{Eq:V}
\end{eqnarray}
where,
\begin{eqnarray}
\tilde V 
=
\left ( 
\begin{array}{ccc}
\tilde c_{12}\tilde c_{13} & \tilde s_{12}\tilde c_{13} & \tilde s_{13} e^{-i\delta_V} \\
- \tilde s_{12}\tilde c_{23} - \tilde c_{12}\tilde s_{23}\tilde s_{13} e^{i\delta_V} & \tilde c_{12}\tilde c_{23} - \tilde s_{12}\tilde s_{23}\tilde s_{13}e^{i\delta_V} & \tilde s_{23}\tilde c_{13} \\
 \tilde s_{12}\tilde s_{23} - \tilde c_{12}\tilde c_{23}\tilde s_{13}e^{i\delta_V} &- \tilde c_{12}\tilde s_{23} - \tilde s_{12}\tilde c_{23}\tilde s_{13}e^{i\delta_V} &  \tilde c_{23}\tilde c_{13} \\
\end{array}
\right).
\end{eqnarray}
with $\tilde c_{ij}=\cos\phi_{ij}$, and $\tilde s_{ij}=\sin\phi_{ij}$.

With magic texture for Majorana-type mass matrix $M_{\rm Mi}$, the Dirac flavor neutrino mass matrix $M_{\rm D}$ can be decompose into 
\begin{eqnarray}
 \label{Eq:eqDeltaMmass}
M_{\rm D}&=&M_{\rm{Mi}}+\Delta M_{\rm{i}}\nonumber \\
&=&
M_{\rm{Mi}}
+
\left(
\begin{array}{ccc}
\ 0 \ & \ 0 \ & \ 0 \ \\
 \delta_{1}& 0 & \delta_{1}^{\rm{Mi}} \\
 \delta_{2} &\delta_{2}^{\rm{Mi}}& \delta_{3}^{\rm{Mi}}\\
\end{array}
\right) 
~~(\rm{i}=1,2,3,4,7,9,10),
\end{eqnarray}
\begin{eqnarray}
\label{Eq:eqDeltaMmass5}
M_{\rm D}&=&M_{\rm{M5}}+\Delta M_{\rm{5}}\nonumber \\
&=&
M_{\rm{M5}}+
\left(
\begin{array}{ccc}
\ 0 \ &\ 0 \ & \ 0 \ \\
 \delta_{1} & \delta_{1}^{\rm{M5}} &\delta_{3}^{\rm{M5}}\\
 \delta_{2} &\delta_{2}^{\rm{M5}}& 0 \\
\end{array}
\right),
\end{eqnarray}
and
\begin{eqnarray}
\label{Eq:eqDeltaMmass68}
M_{\rm D}&=&M_{\rm{Mi}}+\Delta M_{\rm{i}}\nonumber \\
&=&
M_{\rm{Mi}}
+
\left(
\begin{array}{ccc}
\ 0 \ & \ 0 \ & \ 0 \ \\
\delta_{1} & \delta_{1}^{\rm{Mi}} &0 \\
\delta_{2} & \delta_{2}^{\rm{Mi}}& \delta_{3}^{\rm{Mi}} \\
\end{array}
\right) 
~~(\rm{i}=6,8).
\end{eqnarray}

According to Ref.\cite{Zhao2017JHEP}, we define the dimensionless parameters to quantify the strength of the violations of the magic textures feature as follows:
\begin{eqnarray}
 \label{Eq:epsilon12}
\epsilon_1=\frac{d-b}{d+b},\quad \epsilon_2=\frac{g-c}{g+c},
\end{eqnarray}
and
\begin{eqnarray}
 \label{Eq:epsilons}
&&\epsilon_{1}^{\rm{M1}}=\frac{f-(a+c-e)}{f+(a+c-e)},\quad \epsilon_{2}^{\rm{M1}}=\frac{h-(a+c-e)}{h+(a+c-e)},
\quad \epsilon_{3}^{\rm{M1}}=\frac{i-(b-c-e)}{i+(b-c+e)},\nonumber\\
&&\epsilon_{1}^{\rm{M2}}=\frac{f-(a+c-e)}{f+(a+c-e)},\quad \epsilon_{2}^{\rm{M2}}=\frac{h-(a+c-e)}{h+(a+c-e)},
\quad \epsilon_{3}^{\rm{M2}}=\frac{i-(b+c-e)}{i+(b+c-e)},\nonumber\\
&&\epsilon_{1}^{\rm{M3}}=\frac{f-(a-c+e)}{f+(a-c+e)},\quad  \epsilon_{2}^{\rm{M3}}=\frac{h-(a-c+e)}{h+(a-c+e)},
\quad \epsilon_{3}^{\rm{M3}}=\frac{i-(b+c-e)}{i+(b+c-e)},\nonumber\\
&&\epsilon_{1}^{\rm{M4}}=\frac{f-(a-c+e)}{f+(a-c+e)},\quad \epsilon_{2}^{\rm{M4}}=\frac{h-(a-c+e)}{h+(a-c+e)}, \quad \epsilon_{3}^{\rm{M4}}=\frac{i-(b-c+e)}{i+(b-c+e)},\nonumber\\
&&\epsilon_{1}^{\rm{M5}}=\frac{e-(a+b-c)}{e+(a+b-c)},\quad  \epsilon_{2}^{\rm{M5}}=\frac{h-(-b+2c)}{h+(-b+2c)}, \quad \epsilon_{3}^{\rm{M5}}=\frac{f-(a+b-c)}{f+(a+b-c)},\nonumber\\
&&\epsilon_{1}^{\rm{M6}}=\frac{e-(a+b-c)}{e+(a+b-c)},\quad  \epsilon_{2}^{\rm{M6}}=\frac{h-f}{h+f}, \hspace*{1.8cm}\epsilon_{3}^{\rm{M6}}=\frac{i-(a+b-f)}{i+(a+b-f)},\nonumber\\
&&\epsilon_{1}^{\rm{M7}}=\frac{f-(-b+2c)}{f+(-b+2c)},\quad  \epsilon_{2}^{\rm{M7}}=\frac{h-(-b+2c)}{h+(-b+2c)}, \hspace*{0.7cm} \epsilon_{3}^{\rm{M7}}=\frac{i-(b-c+e)}{i+(b-c+e)},\nonumber\\
&&\epsilon_{1}^{\rm{M8}}=\frac{e-(a+b-c)}{e+(a+b-c)},\quad  \epsilon_{2}^{\rm{M8}}=\frac{h-f}{h+f}, \hspace*{1.8cm} \epsilon_{3}^{\rm{M8}}=\frac{i-(-a+2c)}{i+(-a+2c)},\nonumber \\
&&\epsilon_{1}^{\rm{M9}}=\frac{f-(-b+2c)}{f+(-b+2c)},\quad  \epsilon_{2}^{\rm{M9}}=\frac{h-(-b+2c)}{h+(-b+2c)}, \hspace*{0.7cm} \epsilon_{3}^{\rm{M9}}=\frac{i-(-a+2c)}{i+(-a+2c)},\nonumber \\
&&\epsilon_{1}^{\rm{M10}}=\frac{f-(a-c+e)}{f+(a-c+e)},\quad  \epsilon_{2}^{\rm{M10}}=\frac{h-(a-c+e)}{h+(a-c+e)},\quad  \epsilon_{3}^{\rm{M10}}=\frac{i-(-a+2c)}{i+(-a+2c)}.
\end{eqnarray}
By using Eq.(\ref{Eq:epsilon12}) and Eq.(\ref{Eq:epsilons}), $\delta_{i}$ and $\delta_{i}^{\rm{Mi}}$ can be written as
\begin{eqnarray}
 \label{Eq:delta12}
 \delta_{1}=\epsilon_1(b+c), \quad  \delta_{2}=\epsilon_2(g+c),
  \end{eqnarray}
and
\begin{eqnarray}
\label{Eq:deltas}
&&\delta_{1}^{\rm{M1}}=\epsilon_{1}^{\rm{M1}}\{f+(a+c-e)\}, \quad \delta_{2}^{\rm{M1}}=\epsilon_{2}^{\rm{M1}}\{h+(a+c-e)\}, \quad \delta_{3}^{\rm{M1}}=\epsilon_{3}^{\rm{M1}}\{i+(b-c-e)\}, \nonumber \\ 
&&\delta_{1}^{\rm{M2}}=\epsilon_{1}^{\rm{M2}}\{f+(a+c-e)\}, \quad \delta_{2}^{\rm{M2}}=\epsilon_{2}^{\rm{M2}}\{h+(a+c-e)\}, \quad \delta_{3}^{\rm{M2}}=\epsilon_{3}^{\rm{M2}}\{i+(b+c-e)\}, \nonumber \\ 
&&\delta_{1}^{\rm{M3}}=\epsilon_{1}^{\rm{M3}}\{f+(a-c+e)\}, \quad \delta_{2}^{\rm{M3}}=\epsilon_{2}^{\rm{M3}}\{h+(a-c+e)\}, \quad \delta_{3}^{\rm{M3}}=\epsilon_{3}^{\rm{M3}}\{i+(b+c-e)\}, \nonumber \\ 
&&\delta_{1}^{\rm{M4}}=\epsilon_{1}^{\rm{M4}}\{f+(a-c+e)\}, \quad \delta_{2}^{\rm{M4}}=\epsilon_{2}^{\rm{M4}}\{h+(a-c+e)\}, \quad \delta_{3}^{\rm{M4}}=\epsilon_{3}^{\rm{M4}}\{i+(b-c+e)\},\nonumber \\ 
&& \delta_{1}^{\rm{M5}}=\epsilon_{1}^{\rm{M5}}\{e+(a+b-c)\}, \hspace*{0.4cm} \delta_{2}^{\rm{M5}}=\epsilon_{2}^{\rm{M5}}\{h+(-b+2c)\}, \quad~~ \delta_{3}^{\rm{M5}}=\epsilon_{2}^{\rm{M5}}\{f+(a+b-c)\},\nonumber \\ 
 &&\delta_{1}^{\rm{M6}}=\epsilon_{1}^{\rm{M6}}\{e+(a+b-c)\},  \hspace*{0.4cm} \delta_{2}^{\rm{M6}}=\epsilon_{2}^{\rm{M6}}\{h+f\}, \hspace*{1.81cm} \delta_{3}^{\rm{M6}}=\epsilon_{3}^{\rm{M6}}\{i+(a+b-f)\},\nonumber \\ 
 &&\delta_{1}^{\rm{M7}}=\epsilon_{1}^{\rm{M7}}\{f+(-b+2c)\}, \hspace*{0.5cm} \delta_{2}^{\rm{M7}}=\epsilon_{2}^{\rm{M7}}\{h+(-b+2c)\}, \hspace*{0.6cm} \delta_{3}^{\rm{M7}}=\epsilon_{3}^{\rm{M7}}\{i+(b-c+e)\}, \nonumber \\ 
&& \delta_{1}^{\rm{M8}}=\epsilon_{1}^{\rm{M8}}\{e+(a+b-c)\}, \hspace*{0.4cm} \delta_{2}^{\rm{M8}}=\epsilon_{2}^{\rm{M8}}\{h+f\}, \hspace*{1.82cm} \delta_{3}^{\rm{M8}}=\epsilon_{3}^{\rm{M8}}\{i+(-a+2c)\},\nonumber \\ 
&& \delta_{1}^{\rm{M9}}=\epsilon_{1}^{\rm{M9}}\{f+(-b+2c)\}, \hspace*{0.55cm} \delta_{2}^{\rm{M9}}=\epsilon_{2}^{\rm{M9}}\{h+(-b+2c)\}, \hspace*{0.6cm} \delta_{3}^{\rm{M9}}=\epsilon_{3}^{\rm{M9}}\{i+(-a+2c)\},\nonumber \\ 
&&\delta_{1}^{\rm{M10}}=\epsilon_{1}^{\rm{M10}}\{f+(a-c+e)\}, \quad \delta_{2}^{\rm{M10}}=\epsilon_{2}^{\rm{M10}}\{h+(a-c+e)\}, \nonumber \\
&&\delta_{3}^{\rm{M10}}=\epsilon_{3}^{\rm{M10}}\{i+(-a+2c)\}.
\end{eqnarray}
Moreover, we define 
\begin{eqnarray}
|\epsilon|=|\epsilon_{1}|+|\epsilon_{2}|, \quad |\epsilon_{\rm{Mi}}|=|\epsilon_{1}^{\rm{Mi}}|+|\epsilon_{2}^{\rm{Mi}}|+|\epsilon_{3}^{\rm{Mi}}|,
\end{eqnarray}
and 
\begin{eqnarray}
|E_{\rm{Mi}}|=|\epsilon|+|\epsilon_{\rm{Mi}}|.
\end{eqnarray}
If $|E_{\rm{Mi}}|\ll 1$ (in other words $\Delta M_{\rm{i}}$ is small), $M_{\rm{D}}$ is approximately the same as type Mi magic texture.
 
A global analysis of current data shows the following the best-fit values of the squared mass differences $\Delta m_{ij}^2=m_i^2-m_j^2$ and the mixing angles in the case of normal mass ordering (NO), $m_1<m_2<m_3$,\cite{Esteban2019JHEP}:
\begin{eqnarray} 
\frac{\Delta m^2_{21}}{10^{-5} {\rm eV}^2} &=& 7.42^{+0.21}_{-0.20} \quad (6.82 \sim 8.04), \nonumber \\
\frac{\Delta m^2_{31}}{10^{-3}{\rm eV}^2} &=& 2.510^{+0.027}_{-0.027} \quad (2.430 \sim 2.593), \nonumber \\
\theta_{12}/^\circ &=& 33.45^{+0.77}_{-0.75} \quad (31.27 \sim 35.87), \nonumber \\
\theta_{23}/^\circ &=& 42.1^{+1.1}_{-0.9} \quad (39.7 \sim 50.9), \nonumber \\
\theta_{13}/^\circ &=& 8.62^{+0.12}_{-0.12} \quad (8.25 \sim 8.98), \nonumber \\
\delta/^\circ &=& 230^{+36}_{-25} \quad (144 \sim 350), 
\label{Eq:neutrino_observation_NO}
\end{eqnarray}
where the $\pm$ denotes the $1 \sigma$ region and the parentheses denote the $3 \sigma$ region.For the inverted mass ordering (IO), $m_3 < m_1<m_2$, we have
\begin{eqnarray} 
\frac{\Delta m^2_{21}}{10^{-5} {\rm eV}^2}&=& 7.42^{+0.21}_{-0.20} \quad (6.82 \sim 8.04), \nonumber \\
\frac{\Delta m^2_{32}}{10^{-3}{\rm eV}^2} &=& -2.490^{+0.026}_{-0.028} \quad (-2.574 \sim -2.410), \nonumber \\
\theta_{12}/^\circ &=& 33.45^{+0.78}_{-0.75} \quad (31.27 \sim 35.87), \nonumber \\
\theta_{23}/^\circ &=& 49.0^{+0.9}_{-1.3} \quad (39.8 \sim 51.6), \nonumber \\
\theta_{13}/^\circ &=& 8.61^{+0.14}_{-0.12} \quad (8.24 \sim 9.02), \nonumber \\
\delta/^\circ &=& 278^{+22}_{-30} \quad (194  \sim 345).
\label{Eq:neutrino_observation_IO}
\end{eqnarray}
Moreover, we have the following constraints  
\begin{eqnarray} 
\sum m_i < 0.12 - 0.69 ~{\rm eV},
\end{eqnarray}
from the cosmological observation of the cosmic microwave background radiation\cite{Planck2018, Capozzi2020PRD,Giusarma2016PRD,Vagnozzi2017PRD,Giusarma2018PRD}.

In our numerical calculations, we require that the square mass differences $\Delta m_{ij}^{2}$, mixing angles $\theta_{ij}$, and Dirac CP-violating phase $\delta$ should varied with the $3\sigma$ region and the lightest neutrino mass $m_{\rm{lightest}}$ ($m_{\rm{lightest}}=m_1$ for NO and $m_{\rm{lightest}}=m_3$ for IO) is varied within $0-0.12$eV. We also require that the constraints $\sum m_i < 0.12 ~{\rm eV}$ are satisfied.
\begin{table}
 \caption{$|\epsilon|$, $|\epsilon_1|, |\epsilon_2|$, $|\epsilon_{\rm{Mi}}|$, $|\epsilon_{1}^{\rm{Mi}}|$, $|\epsilon_{2}^{\rm{Mi}}|$ and $|\epsilon_{3}^{\rm{Mi}}|$ for the smallest $|E_{\rm{Mi}}|_{\rm min}$.}
 \label{tbl:Epsilons}
 \centering
 \hspace*{-1cm} 
  \begin{tabular}{cccccccccc} 
   \hline
NO& type&$|E_{\rm{Mi}}|_{\rm{min}}$ &$|\epsilon|$ &$|\epsilon_1|$ &$|\epsilon_2|$&$|\epsilon_{\rm{Mi}}|$&$|\epsilon_{1}^{\rm{Mi}}|$&$|\epsilon_{2}^{\rm{Mi}}|$&$|\epsilon_{3}^{\rm{Mi}}|$ \\
\hline
&1&0.461&0.207&0.112&0.0953&0.254&0.0753&0.114&0.0642\\
&2&0.613&0.270&0.0850&0.185&0.343&0.0974&0.121&0.125\\
&3&0.648&0.178&0.0830&0.0948&0.471&0.254&0.143&0.0732\\
&4&0.395&0.0960&0.0769&0.0192&0.299&0.147&0.0373&0.115\\
&5&0.726&0.141&0.0843&0.0563&0.586&0.327&0.197&0.0624\\
&6&0.634&0.365&0.123&0.242&0.269&0.167&0.0363&0.0653\\
&7&0.534&0.365&0.123&0.242&0.169&0.0524&0.0180&0.0986\\
&\bf{8}&\bf{0.377}&\bf{0.147}&\bf{0.0549}&\bf{0.0923}&\bf{0.230}&\bf{0.0696}&\bf{0.156}&\bf{0.00412}\\
&9&0.532&0.221&0.0874&0.134&0.311&0.206&0.00777&0.0968\\
&10&0.545&0.220&0.0720&0.148&0.324&0.0577&0.185&0.0818\\
\hline
\hline
IO& type&$|E_{\rm{Mi}}|_{\rm{min}}$ &$|\epsilon|$ &$|\epsilon_1|$ &$|\epsilon_2|$&$|\epsilon_{\rm{Mi}}|$&$|\epsilon_{1}^{\rm{Mi}}|$&$|\epsilon_{2}^{\rm{Mi}}|$&$|\epsilon_{3}^{\rm{Mi}}|$\\
\hline
&\bf{1}&\bf{0.386}&\bf{0.201}&\bf{0.0940}&\bf{0.107}&\bf{0.185}&\bf{0.108}&\bf{0.0601}&\bf{0.0167}\\
&2&0.720&0.124&0.0709&0.0532&0.596&0.071&0.0734&0.451\\
&3&0.485&0.141&0.0593&0.0814&0.344&0.196&0.0998&0.0481\\
&4&0.502&0.229&0.139&0.0896&0.273&0.177&0.0311&0.0648\\
&5&0.695&0.363&0.144&0.219&0.332&0.120&0.0873&0.124\\
&6&0.602&0.217&0.146&0.0707&0.386&0.127&0.0574&0.201\\
&7&0.532&0.242&0.0661&0.176&0.290&0.0388&0.0858&0.166\\
&8&0.731&0.397&0.240&0.157&0.334&0.165&0.0153&0.153\\
&9&0.887&0.167&0.0642&0.103&0.720&0.277&0.211&0.232\\
&10&0.815&0.357&0.100&0.257&0.459&0.281&0.0952&0.0823\\
\hline
\end{tabular}
\end{table}

Table \ref{tbl:Epsilons} shows the magnitudes of $|\epsilon|$, $|\epsilon_1|, |\epsilon_2|$, $|\epsilon_{\rm{Mi}}|$, $|\epsilon_{1}^{\rm{Mi}}|$, $|\epsilon_{2}^{\rm{Mi}}|$ and $|\epsilon_{3}^{\rm{Mi}}|$ for the smallest $|E_{\rm{Mi}}|_{\rm min}$. The the parameter set for the smallest $|E_{\rm{Mi}}|_{\rm min}$ are indicated by bold fonts for NO and IO cases. In Table \ref{tbl:Epsilons}, we observe 
\begin{itemize}
\item The smallest value of $|E_{\rm{Mi}}|_{\rm min}$ is obtained in the case of type M8 (M1) for the NO (IO).
\item $|\epsilon_2|$ in the type M1 for IO is one order of magnitude greater than that in the type M8 for the NO case.
\item $|\epsilon_{1}^{\rm{M8}}|$ and $|\epsilon_{3}^{\rm{M8}}|$ in the NO case is less than $|\epsilon_{1}^{\rm{M1}}|$ and $|\epsilon_{3}^{\rm{M1}}|$ in the IO case. 
\end{itemize}
Therefore, we can expect that the flavor neutrino mass matrix for Dirac neutrinos approximately obeys type M8 (M1) magic texture for the NO (IO) case. Moreover, since $|E_{M8}|_{\rm{min}}$ for NO is smaller than $|E_{M1}|_{\rm{min}}$ for IO, the NO is slightly preferable to IO in the context of the Magic textures. In addition, comparing $|\epsilon_n|$ and $|\epsilon_{n}^{\rm{M8}}|$ in type M8 for NO case, the $|\epsilon_n|$, $|\epsilon_{1}^{\rm{Mi8}}|$, and $|\epsilon_{3}^{\rm{M8}}|$ are less than $|\epsilon_{2}^{\rm{M8}}|$, suggesting that the violation of type M8 magic texture will be related to $\tau \mu$ mixing. Similarly, comparing $|\epsilon_n|$ and $|\epsilon_{n}^{\rm{M1}}|$ in type M1 for IO case, the $|\epsilon_1|$, $|\epsilon_{2}^{\rm{M1}}|$ and $|\epsilon_{3 }^{\rm{M1}}|$ are less than $|\epsilon_2|$, $|\epsilon_{1}^{\rm{M1}}|$, suggesting that the violation of type M1 magic texture will be related to $\tau e$ and  $\mu \tau$ mixings. 

\begin{table}
 \caption{The neutrino parameters $m_i$ in eV, $\theta_{ij}$ and $\delta$ for the smallest $|E_{\rm{Mi}}|_{\rm min}$.}
 \label{tbl:NPs}
 \centering
 \hspace*{-1cm} 
  \begin{tabular}{cccccccccc} 
   \hline
NO& type&$|E_{\rm{Mi}}|_{\rm{min}}$ & $m_1$ &$m_2$&$m_3$&$\theta_{12}$&$\theta_{23}$&$\theta_{13}$&$\delta$ \\
\hline
&1&0.461&0.00671&0.0110&0.0498&35.5&50.3&8.55&307\\
&2&0.613&0.0238&0.0254&0.0550&35.6&49.8&8.76&163\\
&3&0.648&0.0187&0.0205&0.0531&34.1&47.4&8.84&303\\
&4&0.395&0.0224&0.0240&0.0542&34.0&40.6&8.82&228\\
&5&0.726&0.0236&0.0251&0.0558&32.8&43.5&8.81&300\\
&6&0.634&0.0296&0.0308&0.0575&34.4&46.4&8.65&338\\
&7&0.534&0.0296&0.0308&0.0575&34.4&46.4&8.65&338\\
&\bf{8}&\bf{0.377}&\bf{0.0178}&\bf{0.0197}&\bf{0.0532}&\bf{33.0}&\bf{48.1}&\bf{8.47}&\bf{288}\\
&9&0.532&0.0127&0.0154&0.0514&33.2&48.1&8.62&175\\
&10&0.545&0.0285&0.0297&0.0580&35.3&40.4&8.89&219\\
\hline
\hline
IO& type& $|E_{\rm{Mi}}|_{\rm{min}}$ & $m_1$ &$m_2$&$m_3$&$\theta_{12}$&$\theta_{23}$&$\theta_{13}$&$\delta$\\
\hline
&\bf{1}&\bf{0.386}&\bf{0.0513}&\bf{0.0521}&\bf{0.0120}&\bf{35.1}&\bf{48.5}&\bf{8.75}&\bf{218}\\
&2&0.720&0.0501&0.0509&0.0121&35.1&51.0&8.98&330\\
&3&0.485&0.0499&0.0506&0.0100&31.8&39.8&8.79&295\\
&4&0.502&0.0493&0.0500&0.000524&34.2&43.6&8.35&194\\
&5&0.695&0.0514&0.0521&0.0146&34.0&41.1&8.41&268\\
&6&0.602&0.0503&0.0510&0.00664&34.9&43.3&8.48&302\\
&7&0.532&0.0509&0.0515&0.0127&34.4&40.3&8.79&198\\
&8&0.731&0.0500&0.0508&0.00930&35.9&42.4&9.01&200\\
&9&0.887&0.0495&0.0503&0.00607&33.0&42.8&8.28&329\\
&10&0.815&0.0518&0.0525&0.0146&33.6&45.0&8.26&306\\
\hline
\end{tabular}
\end{table}

Table \ref{tbl:NPs} presents the neutrino parameters $\{m_1,m_2,m_3,\theta_{12},\theta_{23},\theta_{13},\delta\}$ for $|E_{\rm{Mi}}|_{\rm min}$. Regardless the mass ordering, we see that $\theta_{23}>45^\circ$ and $\delta>180^\circ$ are favorable in the context of magic texture. 
There is a tension between T2K and NOvA \cite{T2K2021PRD,NOvA2021arXiv}. Both experiments favor the upper
octant of $\theta_{23}$. However, NOvA and T2K reported very different best fit value of $\delta$ for NO case. In fact, $\delta \sim 145^\circ $ is reported from NOvA \cite{NOvA2021arXiv}. Meanwhile, it is reported as $\delta \sim 250^\circ$ from T2K \cite{T2K2021PRD}. The predicted value $\delta=288^\circ$ for type M8 is near the T2K best-fit value. It seems that the T2K result is more favorable than the NOvA result in the context of the magic textures.
\begin{figure}
\begin{center}
\hspace*{2cm}
\includegraphics{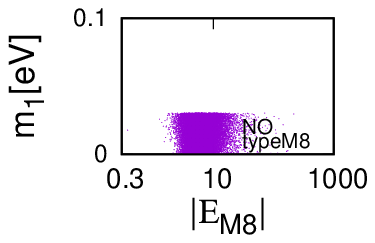} 
\includegraphics{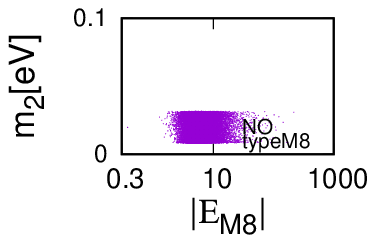}
\includegraphics{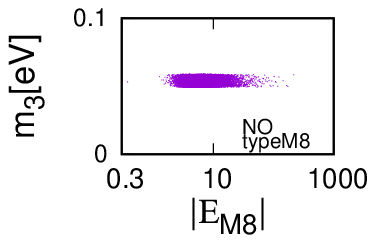}\\
\hspace*{2cm}
\includegraphics{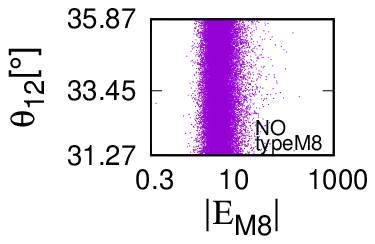}
\includegraphics{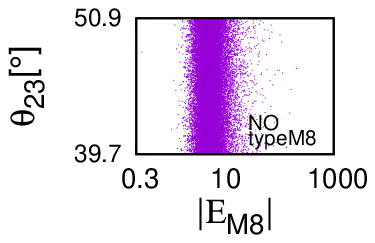}
\includegraphics{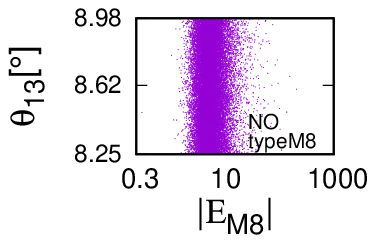}
\hspace*{2.1cm}
\includegraphics{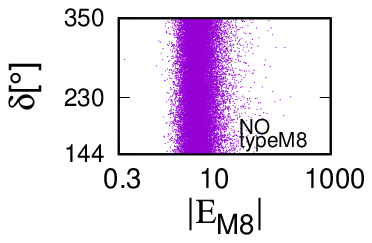}
\vspace{1.5cm}
\caption{Dependence of the neutrino parameters $m_i$ in eV, $\theta_{ij}$, and $\delta$ on $|E_{\rm{M8}}|$ in the case of NO.}
\label{fig:NO_type8}
\end{center}
\end{figure}

\begin{figure}
\begin{center}
\hspace*{2cm}
\includegraphics{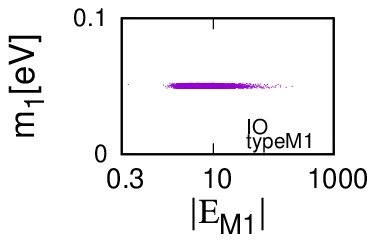} 
\includegraphics{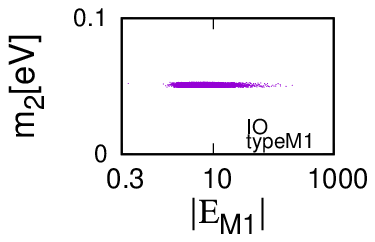}
\includegraphics{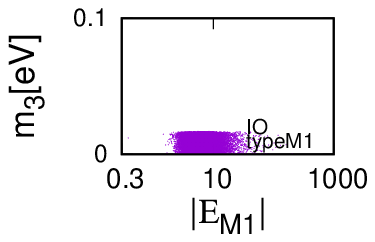}\\
\hspace*{2cm}
\includegraphics{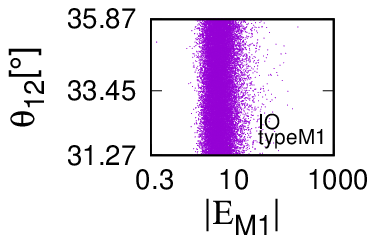}
\includegraphics{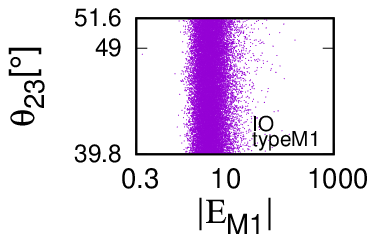}
\includegraphics{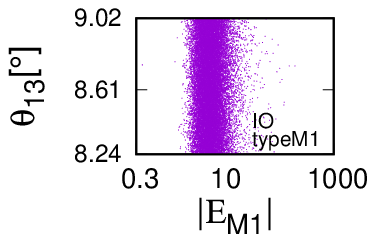}
\hspace*{2.1cm}
\includegraphics{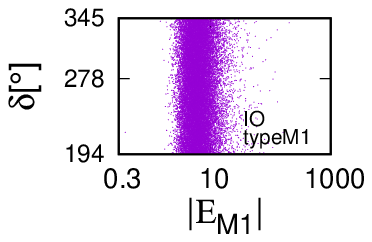}
\vspace{1.5cm}
\caption{Dependence of the neutrino parameters $m_i$ in eV, $\theta_{ij}$, and $\delta$ on $|E_{\rm{M1}}|$ in the case of IO.}
\label{fig:IO_type1}
\end{center}
\end{figure}

Fig.\ref{fig:NO_type8} (Fig.\ref{fig:IO_type1}) shows the dependence of the neutrino parameters $\{m_{i},\theta_{ij},\delta\}$ on the $|E_{\rm{M8}}|$ $(|E_{\rm{M1}}|)$ for the NO (IO) case. The $|E_{\rm{M8}}|$ does not correlate with the neutrino masses $m_1,m_2,m_3$, the mixing angles $\theta_{ij}$ and Dirac CP-violating phase $\delta$. Therefore, the smallest values of $|E_{\rm{M8}}|_{\rm min}$ and $|E_{\rm{M1}}|_{\rm min}$ are obtained for very specific neutrinos parameters.
\section{Summary\label{section:summary}}
The magic textures are one of the successful textures of the flavor neutrino mass matrix for the Majorana neutrinos. 

In this paper, we estimated the relationship between the Dirac flavor neutrino mass matrix and magic texture for Majorana-type neutrinos. We have shown that some type of magic texture for Majorana neutrinos may approximately immanent in the Dirac flavor neutrino mass matrix. More concretely, in the cases of type M8 for NO ($|E_{\rm{M8}}|_{\rm{min}}=0.377$) and type M1 for IO ($|E_{\rm{M1}}|_{\rm{min}}= 0.386$), the Dirac mass matrices are approximately obey the magic texture. Since $|E_{\rm{M8}}|_{\rm{min}}$ for NO is smaller than $|E_{\rm{M8}}|_{\rm{min}}$ for IO, the NO is slightly preferable to IO in the context of the magic textures.
\vspace{1cm}


\begin{thebibliography}{0}    
\bibitem{Hyodo2021PTEP} 
Y. Hyodo and T. Kitabayashi, \Journal{\PTEP}{2021}{123B08}{2021}.

\bibitem{Harrison2004PLB}
P. F. Harrison and W. G. Scott, \Journal{\PLB}{594}{324}{2004}.

\bibitem{Lam2006PLB}
C. S. Lam, \Journal{\PLB}{640}{260}{2006}.

\bibitem{Gautam2016PRD}
R. R. Gautam and S. Kumar, \Journal{\PRD}{94}{036004}{2016}.

\bibitem{Yang2021arXiv}
M. J. S. Yang, \Journal{\PTEP}{2022}{013B12}{2022}.

\bibitem{Channey2019JPGNP}
K. S. Channey and S. Kumar, \Journal{\JPGNP}{46}{015001}{2019}.

\bibitem{Verma2020JPGNP}
S. Verma and M. Kashav,  \Journal{\JPGNP}{47}{085003}{2020}.
\bibitem{Verma2022arXiv} 
M. Kashav, L. Singh and S. Verma arXiv:2207.13328.
\bibitem{Harrison2002PLB}
P. F. Harrison, D. H. Perkins, and W. G. Scott, \Journal{\PLB}{530}{167}{2002}.
\bibitem{Lam2006PRD}
C. S. Lam, \Journal{\PRD}{74}{113004}{2006}.
\bibitem{Hyodo2020IJMPA}
Y. Hyodo and T. Kitabayashi, \Journal{\IJMPA}{35}{2050183}{2020}.
\bibitem{Xing2002PLB}
Z. Z. Xing, \Journal{\PLB}{533}{85}{2002}.

\bibitem{Harrison2002PLB2}
P. F. Harrison and W. G. Scott, \Journal{\PLB}{535}{163}{2002}.

\bibitem{Kitabayashi2007PRD}
T. Kitabayashi, \Journal{\PRD}{76}{033002}{2007}.
\bibitem{Berger2001PRD}
M. S. Berger and K. Siyeon, \Journal{\PRD}{64}{053006}{2001}.

\bibitem{Frampton2002PLB}
P. H. Frampton, S. L. Glashow, and D. Marfatia, \Journal{\PLB}{536}{79}{2002}.

\bibitem{Xing2002PLB530}
Z. Z. Xing, \Journal{\PLB}{530}{159}{2002}.

\bibitem{Xing2002PLB539}
Z. Z. Xing, \Journal{\PLB}{539}{85}{2002}.

\bibitem{Kageyama2002PLB} 
A. Kageyama, S. Kaneko, N. Shimoyana, and M. Tanimoto, \Journal{\PLB}{538}{96}{2002}.

\bibitem{Xing2004PRD}
Z. Z. Xing, \Journal{\PRD}{69}{013006}{2004}.

\bibitem{Grimus2004EPJC}
W. Grimus, A. S. Joshipura, L. Lavoura, and M. Tanimoto, \Journal{\EPJC}{36}{227}{2004}.

\bibitem{Low2004PRD}
C. I. Low, \Journal{\PRD}{70}{073013}{2004}.

\bibitem{Low2005PRD}
C. I. Low, \Journal{\PRD}{71}{073007}{2005}.

\bibitem{Grimus2005JPG}
W. Grimus and L. Lavoura, \Journal{\JPG}{31}{693}{2005}. 

\bibitem{Dev2007PRD}
S. Dev, S. Kumar, S. Verma, and S. Gupta, \Journal{\PRD}{76}{013002}{2007}.

\bibitem{Xing2009PLB}
Z. Z. Xing and S. Zhou,  \Journal{\PLB}{679}{249}{2009}.

\bibitem{Fritzsch2011JHEP}
H. Fritzsch, Z. Z. Xing, and S. Zhou, \Journal{\JHEP}{09}{083}{2011}.

\bibitem{Kumar2011PRD}
S. Kumar, \Journal{\PRD}{84}{077301}{2011}.

\bibitem{Dev2011PLB}
S. Dev, S. Gupta, and R. R. Gautam, \Journal{\PLB}{701}{605}{2011}.

\bibitem{Araki2012JHEP}
T. Araki, J. Heeck, and J. Kubo, \Journal{\JHEP}{07}{083}{2012}.

\bibitem{Ludle2012NPB}
P. Ludle, S. Morisi, and E. Peinado, \Journal{\NPB}{857}{411}{2012}.

\bibitem{Lashin2012PRD}
E. Lashin and N. Chamoun, \Journal{\PRD}{85}{113011}{2012}.

\bibitem{Deepthi2012EPJC}
K. Deepthi, S. Gollu, and R. Mohanta, \Journal{\EPJC}{72}{1888}{2012}.

\bibitem{Meloni2013NPB}
D. Meloni and G. Blankenburg, \Journal{\NPB}{867}{749}{2013}.

\bibitem{Meloni2014PRD}
D. Meloni, A. Meroni, and E. Peinado, \Journal{\PRD}{89}{053009}{2014}.

\bibitem{Dev2014PRD}
S. Dev, R. R. Gautam, L. Singh, and M. Gupta, \Journal{\PRD}{90}{013021}{2014}.

\bibitem{Felipe2014NPB}
R. G. Felipe and H. Serodio, \Journal{\NPB}{886}{75}{2014}.

\bibitem{Ludl2014JHEP}
P. O. Ludl and W. Grimus, \Journal{\JHEP}{07}{090}{2014}.

\bibitem{Cebola2015PRD}
L. M. Cebola, D. E. Costa, and R. G. Felipe, \Journal{\PRD}{92}{025005}{2015}.

\bibitem{Gautam2015PRD}
R. R. Gautam, M. Singh, and M. Gupta, \Journal{\PRD}{92}{013006}{2015}.

\bibitem{Dev2015EPJC}
S. Dev, L. Singh, and D. Raj, \Journal{\EPJC}{75}{394}{2015}.

\bibitem{Kitabayashi2016PRD1}
T. Kitabayashi and M. Yasu\`{e}, \Journal{\PRD}{93}{053012}{2016}.

\bibitem{Zhou2016CPC}
S. Zhou, \Journal{\CPC}{40}{033102}{2016}.

\bibitem{Singh2016PTEP}
M. Singh, G. Ahuja and M. Gupta, \Journal{\PTEP}{2016}{123B08}{2016}.

\bibitem{Kitabayashi2017IJMPA}
T. Kitabayashi, and M. Yasu{\` e}, \Journal{\IJMPA}{32}{1750034}{2017}.

\bibitem{Kitabayashi2017IJMPA2}
T. Kitabayashi, S. Ohkawa and M. Yasu{\` e}, \Journal{\IJMPA}{32}{1750186}{2017}.

\bibitem{Bora2017PRD}
K. Bora, D. Borah and D. Dutta, \Journal{\PRD}{96}{075006}{2017}.

\bibitem{Barreiros2018PRD}
D. M. Barreiros, R. G. Felipe and F. R. Joaquim, \Journal{\PRD}{97}{115016}{2018}.

\bibitem{Kitabayashi2018PRD}
T. Kitabayashi, \Journal{\PRD}{98}{083001}{2018}.

\bibitem{Barreiros2019JHEP}
D. M. Barreiros, R. G. Felipe and F. R. Joaquim, \Journal{\JHEP}{01}{223}{2019}.

\bibitem{Kitabayashi2019IJMPA}
T. Kitabayashi, \Journal{\IJMPA}{34}{1950098}{2019}.

\bibitem{Capozzi2020PRD} 
F. Capozzi, E. D. Valentino and E. Lisi, A. Marrone, A. Melchiorri and A. Palazzo, \Journal{\PRD}{101}{116013}{2020}.

\bibitem{Singh2020EPL}
M. Singh, \Journal{\EPL}{2020}{11002}{2020}.

\bibitem{Barreiros2020}
D. M. Barreiros, F. R. Joaquim and T. T. Yanagida, \Journal{\PRD}{102}{055021}{2020}.

\bibitem{Kitabayashi2020PRD}
T. Kitabayashi, \Journal{\PRD}{102}{075027}{2020}.
\bibitem{Fukuyama1997}
T. Fukuyama and H. Nishiura, (1997), arXiv:hep-ph/9702253.

\bibitem{Lam2001PLB}
C. S. Lam, \Journal{\PLB}{507}{214}{2001}.

\bibitem{Ma2001PRL}
E. Ma and M. Raidal, \Journal{\PRL}{87}{011802}{2001}; Erratum \Journal{\PRL}{87}{159901}{2001}.

\bibitem{Balaji2001PLB}
K. R. S. Balaji, W. Grimus, and T. Schwetz, \Journal{\PLB}{508}{301}{2001}.

\bibitem{Koide2002PRD}
Y. Koide, H. Nishiura, K. Matsuda, T. Kikuchi, and T. Fukuyama, \Journal{\PRD}{66}{093006}{2002}.

\bibitem{Kitabayashi2003PRD}
T. Kitabayashi and M. Yasue, \Journal{\PRD}{67}{015006}{2003}.

\bibitem{Koide2004PRD}
Y. Koide, \Journal{\PRD}{69}{093001}{2004}.

\bibitem{Aizawa2004PRD}
I. Aizawa, M. Ishiguro, T. Kitabayashi, and M. Yasue, \Journal{\PRD}{70}{015011}{2004}.

\bibitem{Ghosal2004MPLA}
A. Ghosal, \Journal{\MPLA}{19}{2579}{2004}.

\bibitem{Mohapatra2005PRD}
R. N. Mohapatra and W. Rodejohann, \Journal{\PRD}{72}{053001}{2005}.

\bibitem{Koide2005PLB}
Y. Koide, \Journal{\PLB}{607}{123}{2005}.

\bibitem{Kitabayashi2005PLB}
T. Kitabayashi and M. Yasue, \Journal{\PLB}{621}{133}{2005}.

\bibitem{Haba206PRD}
N. Haba and W. Rodejohann, \Journal{\PRD}{74}{017701}{2006}.

\bibitem{Xing2006PLB}
Z. Z. Xing, H. Zhang, and S. Zhou, \Journal{\PLB}{641}{189}{2006}.

\bibitem{Ahn2006PRD}
Y. H. Ahn, S. K. Kang, C. S. Kim, and J. Lee, \Journal{\PRD}{73}{093005}{2006}.

\bibitem{Joshipura2008EPJC}
A. S. Joshipura, \Journal{\EPJC}{53}{77}{2008}.

\bibitem{Gomez-Izquierdo2010PRD}
J. C. Gomez-Izquierdo and A. Perez-Lorenzana, \Journal{\PRD}{82}{033008}{2010}.

\bibitem{He2001PRD}
H. J. He and F. R. Yin, \Journal{\PRD}{84}{033009}{2011}.

\bibitem{He2012PRD}
H. J. He and X. J. Xu, \Journal{\PRD}{86}{111301}{2012}.

\bibitem{Gomez-Izquierdo2017EPJC}
J. C. Gomez-Izquierdo, \Journal{\EPJC}{77}{551}{2017}.

\bibitem{Fukuyama2017PTEP}
T. Fukuyama, \Journal{\PTEP}{2017}{033B11}{2017}.

\bibitem{Kitabayashi2016IJMPA}
T. Kitabayashi, \Journal{\IJMPA}{31}{09}{2016}.

\bibitem{Kitabayashi2016PRD}
T. Kitabayashi, and M. Yasu{\` e}, \Journal{\PRD}{94}{075020}{2016}.

\bibitem{Bao2021arXiv}
Z. H. Zhao, X. Y. Zhao, and H. C. Bao, \Journal{\PRD}{105}{035011}{2022}.
\bibitem{Garces2018JHEP} 
E. A. Garc{\' e}s, Juan Carlos G{\' o}mez-Izquierdo and F. Gonzalez-Canales \Journal{\EPJC}{78}{812}{2018}.
\bibitem{JuanCarlos2019JHEP} 
Juan Carlos G{\' o}mez-Izquierdo and Myriam Mondrag{\' o}n \Journal{\EPJC}{79}{285}{2019}.
\bibitem{Altarelli2010PMP}
G. Altarelli and F. Feruglio, \Journal{\RMP}{82}{2701}{2010}.
\bibitem{Borgohain2021JPG} 
H. Borgohain, D. Borah, \Journal{\JPGNP}{48}{075005}{2021}.
\bibitem{Pontecorvo1957}
B. Pontecorvo, Sov. Phys. JETP 6 (1957) 429.

\bibitem{Pontecorvo1958}
B. Pontecorvo, Sov. Phys. JETP 7 (1958) 172;

\bibitem{Maki1962PTP}
Z. Maki, M. Nakagawa and S. Sakata, \Journal{\PTP}{28}{870}{1962}.

\bibitem{PDG}
M. Tanabashi {\it et al.} (Particle Data Group), \Journal{\PRD}{98}{030001}{2018}.
\bibitem{Senjanovic2016PRD} 
G. Senjanovi{\' c}, and V. Tello, \Journal{\PRD}{94}{095023}{2016}.
\bibitem{Zhao2017JHEP} 
Z. Zhao, \Journal{\JHEP}{09}{023}{2017}.
\bibitem{Esteban2019JHEP}
M. C. Gonzalez-Garcia, Michele Maltoni, and Thomas Schwetz, Universe 2021, 7(12),
459(2021), “ NuFit 5.1 (2021) ”, www.nu-fit.org.

\bibitem{Planck2018}
N. Aghanim, et al. (Planck Collaboration), ``Planck 2018 results. VI. Cosmological parameters", arXiv:1807.06209.

\bibitem{Giusarma2016PRD}
E. Giusarma, M. Gerbino, O. Mena, S. Vagnozzi, S. Ho and K. Freese, \Journal{\PRD}{94}{083522}{2016}.

\bibitem{Vagnozzi2017PRD}
S. Vagnozzi, E. Giusarma, O. Mena, K. Freese,  M. Gerbino, S. Ho and M. Lattanzi, \Journal{\PRD}{96}{123503}{2017}.

\bibitem{Giusarma2018PRD}
E. Giusarma, S. Vagnozzi, S. Ho, S. Ferraro, K. Freese,  R. K. Rubio and K. B. Luk, \Journal{\PRD}{98}{123536}{2018}.
\bibitem{NOvA2021arXiv}
M. A. Acero, et al. (NOvA Collaboration), \Journal{\PRD}{106}{032004}{2022}.

\bibitem{T2K2021PRD}
K. Abe, et al. (T2K Collaboration), \Journal{\PRD}{103}{112008}{2021}.
\end{thebibliography}
\end{document}